\documentclass[prl,twocolumn,times,showpacs,superscriptaddress]{revtex4-1}
\usepackage{amsmath, amssymb, braket, dsfont}
\usepackage[mathscr]{euscript}
\usepackage{graphicx}
\usepackage{subfigure}		
\usepackage{epstopdf, times}
\usepackage[breaklinks]{hyperref}

\usepackage[usenames,dvipsnames]{color}			

\newcommand{\T}{{\mathcal T}}
\newcommand{\gsd}{\mathfrak{m}}
\newcommand{\F}{\mathcal{F}}
\begin{document}

\title{Constraints on topological order in Mott Insulators}
\author{Michael P. Zaletel}
\affiliation{Department of Physics, University of California, Berkeley, California 94720, USA}
\affiliation{Department of Physics, Stanford University, Stanford, California 94305, USA}
\author{Ashvin Vishwanath}
\affiliation{Department of Physics, University of California, Berkeley, California 94720, USA}

\begin{abstract}
We point out certain symmetry induced constraints on topological order in Mott Insulators (quantum magnets with an odd number of spin $\tfrac{1}{2}$ per unit cell). We show, for example, that the double semion topological order is incompatible with time reversal and translation symmetry in Mott insulators. This sharpens the Hastings-Oshikawa-Lieb-Schultz-Mattis theorem for 2D quantum magnets, which guarantees that a fully symmetric gapped Mott insulator must be topologically ordered, but is silent on which topological order is permitted. An application of our result is the Kagome lattice quantum antiferromagnet where recent numerical calculations of entanglement entropy indicate a ground state compatible with either toric code or double-semion topological order. Our result rules out the latter possibility.
\end{abstract}

\maketitle

Distinctions between phases of matter were traditionally based on symmetry considerations, since spontaneous symmetry breaking leads to distinct phases.
However, with the discovery of the fractional Quantum Hall effects (FQHE), the role of topology in defining phases of matter was emphasized.
Topologically ordered states in two dimensions (2D), such as FQHE phases and gapped quantum spin liquids, contain excitations with unusual (anyonic) statistics.
Symmetries still have an important role to play in these systems, as the anyon excitations may carry fractional quantum numbers such as the fractional charge of Laughlin quasiparticles.  
The interplay of topology and symmetry leads to new and fundamental distinctions between states of matter. \cite{WenBook}

Charge conservation also allows us to define the filling $\nu$ of FQHE states.
At fractional filling factors $p/q$, a featureless (translation invariant)  state \emph{must} have a non-trivial excitation of charge $1/q$.
This is the simplest example of a constraint between the microscopic details (the fractional filling) and the emergent excitations (the fractional charge).
Presumably this constraint helps stabilize the FQHE states over competing conventional orders which must break translation symmetry.

Gapped quantum spin liquids are close analogs of  FQHE states.
They also feature emergent anyon excitations and fractionalization of symmetry quantum numbers, although usually in the presence of time reversal symmetry.
They are proposed to occur in two dimensional insulating quantum magnets, where frustration prevents the formation of a conventional ordered state\cite{Balents2010}.
Although a clearcut experimental example of a gapped spin liquid is currently lacking, numerical calculations have made a strong case for their existence in the $S=\tfrac{1}{2}$ anti-ferromagnet on the Kagome lattice\cite{Yan2011, Depenbrock2012}.
Experiments on the Kagome lattice material Herbertsmithite also observe a spin disordered state\cite{Helton2007,Helton2010}. Although bulk measurements do not observe an energy gap, this distinction has been attributed to disorder \cite{Harrison2008} or non-Heisenberg magnetic interactions\cite{Mendels2008,Cepas2008,Sachdev2010}, although other ground states have also been proposed\cite{Ran2007}.

The analog of fractional filling in quantum magnets is the Mott insulator, defined as an insulator with an odd number of $S=\tfrac{1}{2}$ moments per unit cell.
In 1D, according to the Lieb-Schultz-Mattis argument, a $S=\tfrac{1}{2}$ antiferromagnetic chain must ether be gapless or double the unit cell\cite{LSM1961}.
In 2D an analog of this result, the Hastings-Oshikawa-Lieb-Schultz-Mattis (HOLSM) \cite{Hastings2005, Oshikawa2000, Paramekanti2004,Auerbach1994B, Parameswaran2013}, states that at zero temperature, a Mott insulator must either be gapless, break spin / translation symmetry, or have emergent excitations with nontrivial mutual statistics.
The last condition is not available in 1D, and corresponds to a topological  quantum spin liquid phase, which is gapped and preserves all symmetries.
Hence finding  a symmetric, gapped state is indirect, but strong, evidence for a quantum spin liquid. 

An intuitive way of visualizing this result is to think of a $S = \tfrac{1}{2}$ in terms of hard core bosons, where spin up is an empty site and spin down is a site occupied by a boson.
A Mott insulator has a fractional (half odd integer) filling of bosons per unit cell.
To obtain a featureless insulator, the bosons must fractionalize into half charged entities, which can then be uniformly assigned to lattice sites.
When viewed directly from the spin language, this implies that to obtain a symmetric ground state, one needs $S=\tfrac{1}{2}$ excitations in the magnet which can screen the background spin in the unit cell.  
No local excitation (like a spin flip) carries $S = \tfrac{1}{2}$, so these excitations must be topological.

Clearly this will place conditions on the types of topological order compatible with a symmetric state.
The extensions of the Lieb-Schultz-Mattis theorem are silent on the detailed form of the topological order.
Here we will  show that one very natural seeming type of topological order, the double-semion state, is incompatible with a time reversal symmetric Mott insulator.
Our method of proof can be readily generalized to other kinds of topological order and different symmetries.
We leave that to future work.

This observation has an important consequence for interpreting recent numerical results on the Kagome antiferromagnet.
Numerical calculations using the Density Matrix Renormalization Group have found a gapped phase with a featureless ground state i.e. one that preserves the spin, lattice and time reversal symmetry\cite{Yan2011}.
As the Kagome model is a Mott insulator with three $S=\tfrac{1}{2}$ per unit cell, this implies a quantum spin liquid phase.
Subsequently, the topological entanglement entropy  was calculated in this ground state and was found to be consistent with $\gamma=\ln 2$,\cite{Jiang2012,Depenbrock2012}  the expected value for a spin liquid with Z$_2$ toric code topological order.
Certain other topological orders are also  compatible with this value but they break time reversal symmetry.
The only  other plausible option is the double-semion theory, which is a twisted Z$_2$ topological order \cite{Freedman2004, LevinWen}.
Excitations in this phase are a semion, antisemion and a boson with mutual semionic statistics with the first two particles.
Our argument demonstrates that double-semion topological order is incompatible with a fully symmetric Mott insulator.
Given that the numerical results on the Kagome lattice antiferromagnet point to a symmetric ground state, we can exclude this topological order.
The only remaining possibility which is consistent with all numerical results is the Z$_2$ toric code topological order. 

This paper is organized as follows.
First, we provide a summary of the central result and its application to the double-semion theory.
We then give a more rigorous argument based on the action of symmetries on the minimally entangled ground states of an infinite cylinder.
Finally, we discuss some numerical studies of ground states with double semion topological order in a  Kagome Mott insulator,  which provide a concrete illustration of the constraints described in this paper.

We will argue that in the presence of translation symmetry, there is  an Abelian anyon $a$ in the system which \textbf{1}) is not transformed into another anyon type under the symmetries and \textbf{2}) can `screen' the charge in the unit cell, meaning that $a$ transforms under the symmetries in a manner  that can combine with a missing unit cell in order to form a neutral object.
One may visualize the system as a lattice of $a$-particles which screen the fractional charge of the unit cell. 
We then show that there are no anyons in the  double-semion theory which satisfy criteria \textbf{1}, \textbf{2}), completing a `no-go' argument.

There is an exception to our argument for certain exotic realizations of translation, such as when translation permutes the anyon types. This is impossible for the double-semion theory, so we defer all discussion of this case to the Supplementary Materials.  \cite{SuppMat}

The properties of a topological phase with respect to translation \cite{Wen2002, Wen2003, Kitaev2006, EssinHermele2013, FidkowskiLindnerKitaev, Wang2014} can be captured by supposing there is an Abelian anyon of type `$a$' sitting in each unit cell, generating a constant density of topological flux $a$. 
The  anyons then experience  magnetic flux owing to their mutual statistics with the anyon $a$. 
To be precise,  \cite{SuppMat} take an anyon $b$ around a  path enclosing one unit cell, accumulating a Berry phase 
\begin{equation}
\eta_b = \frac{(T^{-1}_y T^{-1}_x T_y T_x)_b} { (T^{-1}_y T_y T^{-1}_x T_x)_b} = S_{b a} /S_{b \mathds{1}},
\label{eq:eta}
\end{equation} 
where $x, y$ denote a basis for the Bravais lattice.
The denominator has been included so that if the state is translationally symmetric, the non-universal components of each $T_{x/y}$ cancel, resulting in a robust phase.
If there is constant topological flux $a$,  then $b$ has enclosed one $a$-flux, accumulating mutual statistics $S_{b a} /S_{b \mathds{1}}$ where $S$ is the topological $S$-matrix.
On physical grounds,  these phases should be consistent with fusion, $\eta_b \eta_c = \eta_{b c}$ (in a non-Abelian phase, all fusion channels for $bc$ should share the same $\eta$). 
Setting $\eta_b = S_{b a} /S_{b \mathds{1}}  $ for some Abelian $a$ is in fact the unique choice consistent with fusion, so measuring each $\eta_b$  uniquely determines (and defines) the flux $a$.

For example, consider the  FQHE at $\nu = 1/m$; the anyons are labeled by their charge $Q_b = e b / m$.
When an anyon $b$ encircles one magnetic unit cell it acquires an Aharanov-Bohm phase $\eta_b = e^{2 \pi i \tfrac{b}{m} }$.
Since $S_{ba} = e^{2 \pi i  b a / m} / \sqrt{m} $, we see that the background topological flux is $a = 1$, the $Q_a = e/m$ quasi-particle.

In the presence of other symmetries, there are two constraints on the allowed background flux $a$.
First (\textbf{1}),  note that in general applying a global symmetry $G$ can turn one anyon type into another, $G: b \to G b$.
The flux $a$ must be left invariant under any symmetry $G$ which commutes with the translations $T_{x/y}$,  otherwise the phases $\eta_b$ will break the symmetry $G$.

Second (\textbf{2}), we will later prove that the anyon $a$ transforms under the symmetries so as to screen the microscopic unit cell.
For concreteness we discuss three cases, each of which is applicable to the Mott insulator:
i) if there is half-integral spin per unit cell,  $a$ must have half-integral spin (it is a `spinon');
ii) if there is fractional U(1) charge $n/m$ per unit cell, then $a$ must have fractional charge $n/m$ (as for the Laughlin quasiparticles);
and iii) if each unit cell transforms as a Kramer's doublet $\T^2 = -1$, then $a$ must transform as a Kramer's doublet.

Cases i - iii imply there must be non-trivial topological order: since the charges assigned to $a$ are fractional, they cannot be carried by any local (trivial) excitation, which is the content of the HOLSM theorem.
But from (\textbf{1}) we have learned something in addition to HOLSM: $a$ cannot be permuted by the symmetries.
This small addition is sufficient to rule out the double-semion theory.

The double-semion topological order can be viewed as a topological phase of bosons which is comprised of a pair of opposite $m=\pm2$ bosonic Laughlin states ($U(1)_2\times U(1)_{-2}$).
It can be described by a two component Abelian Chern Simons theory $ {\mathcal L} = \frac2{4\pi}\epsilon^{\mu\nu\lambda}(a_{1\mu}\partial_\nu a_{1\lambda}- a_{2\mu}\partial_\nu a_{2\lambda})$ and has the same quantum dimension and ground state degeneracy on the torus (4) as the Z$_2$ toric code topological order.
The quasiparticle content is $\{1,\,s\} \times \{1,\,s'\}=\{1,\,s,\,s',\,b\}$ where $s$ ($s'$) is the semion (antisemion) and $b=ss'$ is a boson, with mutual semionic statistics with the first two particles. 

The topological spins of the semions are $\theta_{s/s'} = \pm i$.
Under time reversal, the topological spin is conjugated, $\theta_{\T s} = \theta^\ast_{s}$, so time reversal  exchanges the semions: $\T s = s'$.
This  constrains the allowed realizations of SO(3), U(1) and time reversal symmetry in a way  we show is incompatible with scenarios i-iii).
In all cases we assume that both time reversal and translation symmetry are respected.

\paragraph{i) SO(3).}
	There are two ways to realize SO(3) in the double-semion model.
First, we can assign trivial (integral) spin to each anyon. 
But for case i) we need at least one anyon to transform as $S = \tfrac{1}{2}$, and the unique possibility is that $\{1, b\}$  have integral spin while $\{ s, s'\}$ have half-integral spin.
Clearly  $s, s'$ must have the same spin, as they are related by time reversal.
Since $b s = s'$, $b$ cannot have half-integral spin in order to preserve consistency with fusion.
There is no anyon $a$ which carries $S = \tfrac{1}{2}$ and isn't permuted by $\T$ .

\paragraph{ii) U(1).}	
	Since $b^2 = s^2 = {s'}^2 = 1$, fusion requires that each anyon either has U(1) charge $Q$ of $0$ or $\tfrac{q}{2}$ (modulo the unit of charge $q$ in this theory). 
Fusion and time-reversal require $Q_s = Q_{s'} = Q_{s} + Q_b$, so $Q_b = 0$ is neutral.
There are two possibilities: $Q_{s / s'} = 0$, or $Q_{s / s'} = \tfrac{q}{2}$.
In either case, there is no anyon $a$ which carries  $Q = \tfrac{n}{m} q$  and is not permuted by $\T$.

\paragraph{iii) Time reversal.}
	Under $\T$ the two semions are exchanged, $\T: s \leftrightarrow s'$, while the boson $b$ is unchanged.
Furthermore,  the boson $b$ must be assigned $\T^2 = 1$ because it is composed of a pair of particles ($s,\,s'$) with trivial mutual statistics which  are transformed into one another under time reversal.
Leaving a detailed argument  to the Supplementary Materials \cite{SuppMat}, intuitively  when $\T^2$ acts on $b$ it is equivalent to taking $s$ around $s'$, which leads to unit phase i.e. $\T^2=+1$, and hence no Kramers degeneracy. 
Again, there is no anyon $a$ which is a Kramer's doublet.	
 
To justify criteria \textbf{1}, \textbf{2} we consider  the action of braiding,  time reversal, and translation  on the degenerate ground states of an infinitely long cylinder.
There is a special `minimally entangled' (ME) basis  \cite{Zhang2012} for the degenerate ground states in which each of these operations permutes the basis states.
These permutations are subject to certain conditions which impose the two constraints  \textbf{1}, \textbf{2} on the background anyon flux $a$.
We restrict to Abelian theories to simplify the discussion, but the result is general.

\begin{figure}[t]
\begin{centering}
	\includegraphics[width=\columnwidth]{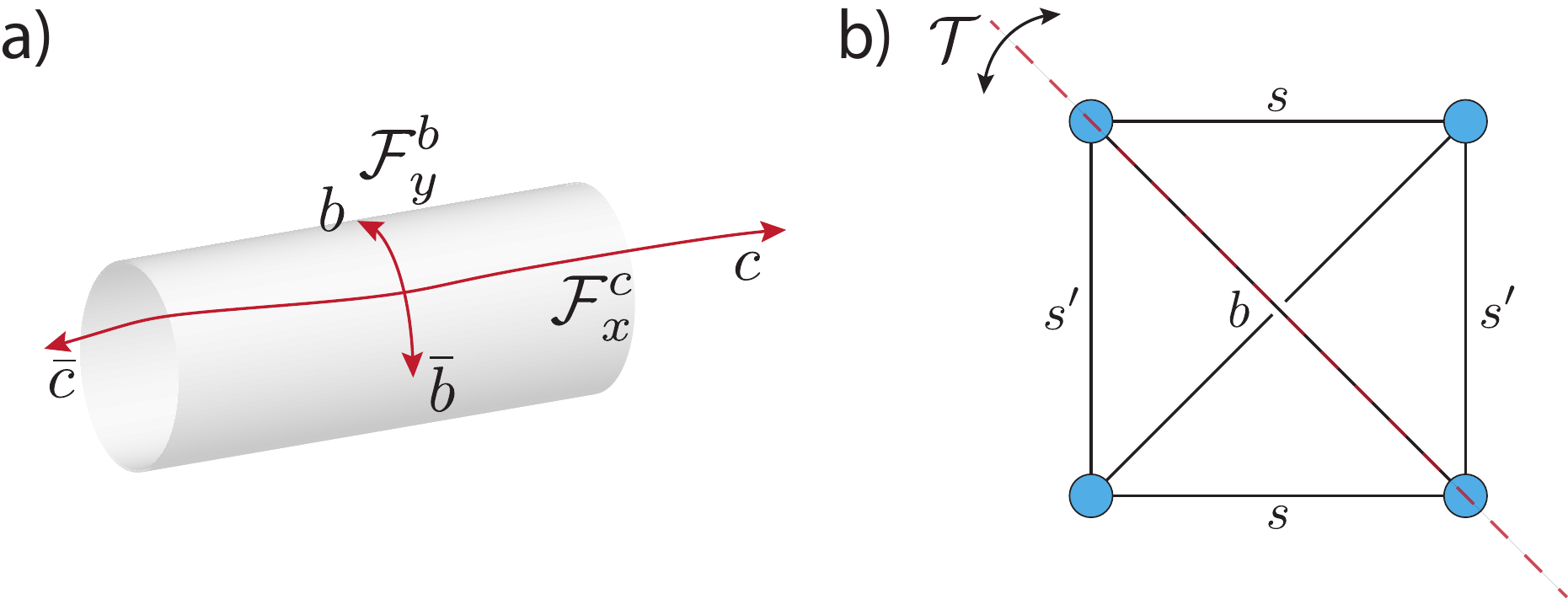}
	\caption{ \textbf{a)} The adiabatic processes $\F_{x/y}^b$.  \textbf{b)} The 4 minimally entangled basis states are represented as the node of a graph.
The process $\F_x^c$, with $c = {1, b, s, s'}$, permutes these basis state, illustrated with labeled edges.
The action of time reversal $\T$ is also a permutation of the MES; the only permutation consistent with fusion acts as a mirror reflection across the diagonal, since it must exchange $\T: s \leftrightarrow s'$}
	\label{fig:mes}
\end{centering}
\end{figure}

A topological theory with  $\gsd$ anyon types has $\gsd$ degenerate ground states  on an infinitely long cylinder\cite{Wen1990}.
To construct the ME basis, \cite{Zhang2012} define periodic coordinate $y$ and infinite coordinate $x$.
Let $\F_y^a$ denote the adiabatic process of creating a pair of anyons $b / \bar{b}$ from the vacuum, taking $b$ around the cylinder in the $+y$ direction, and reannihilating  the pair, as illustrated in Fig.~\ref{fig:mes} \cite{WenNiu1990, OshikawaSenthil2006}. $\F_x^b$ is a similar process in which a pair $b / \bar{b}$ is dragged out to $x  = \pm \infty$.
The $\F_{x/y}^b$ are a set of unitary matrices acting on the ground state manifold.
$\F_x^b$ threads topological flux $b$ through the cylinder, while the $\F_y^b$ are like Wilson loops which detect the topological flux. 
Their commutation relations are determined by the mutual statistics $S_{bc}/S_{b\mathds{1}}$ and the fusion group $N_{bc}^d$:
\begin{align}
\F_{y}^b \F_{x}^c &= \frac{S_{bc}}{S_{b \mathds{1}}} \F_{x}^c   \F_{y}^b: \nonumber \\
\F_{x/y}^b \F_{x/y}^c &\propto \F_{x/y}^{b\cdot c},    \quad b \cdot c = \sum_d N_{b c}^d d \quad \mbox{ (Abelian fusion)}
\label{eq:T_relations}
\end{align}

	The ME basis  simultaneously diagonalizes each $\F^b_{y}$. \cite{KitaevPreskill2006, WenWang2008, Zhang2012}
The ME basis has definite  topological flux threading the cylinder,  reducing the entanglement entropy between the two regions $x < 0$ and $x > 0$.
In contrast, for the non-MES, the Wilson loop - Wilson loop correlation functions generated by $\F_y^b$ have long-range order along the length of the cylinder,  generating additional entanglement entropy ( they are long-range ordered `cat states' if we view the cylinder as a 1D system).
By choosing a basis which diagonalizes $\F^b_{y}$, each basis state is a local minima of the entanglement.

For Abelian $b$, the process $\F_x^b$  \emph{permutes} the MES in a manner consistent with  fusion.
We represent this permutation as a graph, shown in Fig.~\ref{fig:mes} for the double-semion theory.
Each node of the graph is an MES; nodes are connected by an edge `$b$' if the two MES are related by $\F_x^b$.


Time-reversal or an onsite symmetry (such as spin rotations)  $G$ must also \emph{permute} the MES: these symmetries leave the  entanglement entropy invariant, so under $G$ the MES  remain local minima of the entanglement entropy.
When $G$ acts on an anyon, it can also be transformed into some other anyon, $G: b \to G b$.
Since $G \F_y^b G^{-1} \propto \F_y^{G b}$  while  $\F_y^b \F_y^c \propto \F_y^{b\cdot c}$, there are constraints on the allowed permutations of the MES.

As an example, consider time-reversal $\T$ in the double-semion model, where $\T$ leaves the anyons $\mathds{1}, b$ invariant, but  exchanges the semions, $\T : s \leftrightarrow s'$.
Referring to Fig.~\ref{fig:mes}, we see that the permutation $\T$ must act like a reflection across the diagonal, exchanging $s$ edges and $s'$.

Finally we consider translations $T_x$ along the length of the cylinder,  taking an entanglement point of view on the LSM theorem.
Again, $T_x$ can only permute the MES, because the MES are the unique basis states which are not long-range correlated along the length $x$ of the cylinder, and $T_x$ cannot generate long-range correlations.
In fact, $T_x$  is equivalent to threading  topological flux $\F_x^{a^{L_y}}$, where $a$ is the anyon in each unit cell and $L_y$ is the circumference of the cylinder, because $T_x$ transfers $L_y$ of the $a$ through the cylinder.
The commutator ${\F_y^b}^{-1} T^{-1}_x \F_y^b T_x$ is equivalent to an anyon $b$ encircling an annular region of $1 \times L_y$ unit cells. 
As discussed, the result is a robust phase $\eta_b^{L_y}$.
Using Eq.\eqref{eq:eta}, $\eta_b = S_{b a} /S_{b \mathds{1}}$, combined with Eq. \eqref{eq:T_relations} and the non-degeneracy of braiding, we find $T_x \propto \F_x^{a^{L_y}}$.

	To understand the further constraints on the permutation  $T_x$ (and hence on the special anyon $a$) we examine the entanglement properties for bipartitions at different $x$.
Let $\rho_{x}$ be the reduced density matrix for the system to the left of $x$ (leaving the dependence on the particular MES  implicit).
If the state is symmetric (under assumptions  satisfied by the double-semion theory, there is always at least MES which is symmetric \cite{SuppMat}), then under $\T$ or an SO(3) spin rotation $R$, $\rho_x$ transforms as
\begin{align}
\T : \rho_{x}  &\to   U_{\T ; x} \rho^\ast_{x} U_{\T ; x}^{\dagger}, \quad \quad R : \rho_{x}  &\to   U_{R; x} \rho_{x} U_{R; x}^{\dagger}
\end{align}
where $U_{\T ; x}, U_{R; x}$ are unitary matrices. 
It is known that the $U$  are a \emph{projective} representation of the symmetries.  \cite{PollmannTurnerBergOshikawa2010, FidkowskiKitaev2011, ChenGuWen2011}
For  $\T $ there are two possibilities,
\begin{align}
U_{\T ; x} U^\ast_{\T ; x} = \gamma_x,   \quad   \gamma_x = \pm 1
\end{align}
independent of whether the microscopic degrees of freedom transform as $\T ^2 = \pm 1$.
For rotations $R$, the $U_{R; x}$ can either be decomposed into integral representations of SO(3), which we denote by $S_x = 1$, of half-integral representation of SO(3), which we denote $S_x= - 1$.

	We  make use of the odd number of $S = \tfrac{1}{2}$ per unit cell by calculating the dependence of $\gamma_x, S_x$ on the location of the cut $x$.
Consider a cylinder with odd, but arbitrarily large, circumference, so that each ring of the cylinder transforms as $\T ^2 = -1$ and with half-integral spin.
Since the reduced density matrices for $\rho_{x}, \rho_{x+1}$ differ by the addition of a single ring, it is straightforward to prove \cite{ChenGuWen2011} that
\begin{align}
\gamma_{x+1} = - \gamma_x, \quad \quad S_{x+1} = - S_x.
\end{align}
Intuitively, every time the entanglement cut passes a spin, the entanglement invariants flip, since the spins transform with $\gamma = S = -1$ themselves. 
However, the cuts at $x$, $x+1$ are related by the translation $T_x$, so the state must double the unit cell -  this is a version of the LSM theorem.
A similar phenomena occurs whenever the unit-cell transforms projectively under a symmetry.\cite{ChenGuWen2011}
The case of U(1) at fractional filling is somewhat distinct, but the conclusion equivalent \cite{SuppMat}.

Tying these strings together, we argued that $T_x$ is a permutation equivalent to threading some Abelian  flux, $T_x \propto {\F}_x^{a^{L_y}}$.  
For odd $L_y$, $T_x$ flips the entanglement invariants, so $a$ must be non-trivial. 
When an anyon $a$ crosses an entanglement cut during the process  $\F_x^{a}$, the entanglement invariants $\gamma_x / S_x$ flip \emph{if and only if} $a$ transforms as $\T^2 = -1$ / with half-integral spin. 
More generally, we  conclude that $a$ must transform with the same projective representation or $U(1)$ fractional charge as the unit cell; this is the precise meaning of criteria \textbf{2}), that $a$ can `screen' the charge of the unit cell. Criteria \textbf{1}) follows from $T_x \mathcal{T} = \mathcal{T} T_x$.

	Returning to the double-semion model, examining Fig.\ref{fig:mes} we see that the only non-trivial choice consistent with time-reversal is $T_x \propto {\F}_x^{b^{L_y}}$.
But  the bosonic excitation must have integral spin and $\T ^2 = 1$, so when a boson $b$ passes an entanglement cut at $x$ it does \emph{not} flip the entanglement invariants $\gamma_x, S_x$.
But if $\F_x^b$ leaves these invariants unchanged, while   $T_x$  flips them, we arrive at a contradiction.


	Several recent works have examined the possibility of double-semion quantum spin liquids on lattices including the Kagome model.
These works were partially motivated by numerical evidence that there is a chiral spin liquid adjacent to the $S=\tfrac{1}{2}$ Kagome Heisenberg anti-ferromagnetic phase, with tentative evidence that the two phases may be related by a continuous transition. \cite{Bauer2014, HeChenYan2014, He2014, Gong2014} 
There is a natural scenario for a continuous phase transition between a double-semion theory and a chiral spin-liquid. \cite{Barkeshli2013}

	These theoretical studies found exactly solvable quantum-dimer models with double-semion topological order.\cite{QiGuYao2014, IqbalPoilblancSchuch, BuerschaperMorampudiPollmann2014}
In the dimer picture, these double-semion states preserve translation, time reversal, and SO(3).
But this is not a counter example to our no-go argument, because  the dimer picture  loses track of the $S=\tfrac{1}{2}$ nature of the constituent spins.

In fact, Ref.~\onlinecite{IqbalPoilblancSchuch} provides intriguing evidence for the no-go argument.
A dimer wavefunction can be translated into a $S=\tfrac{1}{2}$  wavefunction, but this requires choosing a particular dimer reference configuration.
While the reference configuration breaks translation invariance, when this procedure is applied to the RVB state with the topological order of the Z$_2$ toric-code, the resulting state is translation invariant.
However, when applied to the double-semion RVB, there is an \emph{observable} doubling of the unit cell which could not be removed within the variational space considered.
In light of the no-go argument it appears this is  an intrinsic feature of the $S=\tfrac{1}{2}$ Kagome model.

In conclusion, we have argued that symmetries enforce a new type of constraint on the  topological order of a Mott insulator. Like Lieb-Schultz-Mattis and its extensions, this result is a helpful ally in the hunt for spin liquids since local order parameters cannot be used to distinguish between topological orders.
	
\acknowledgements
MZ is indebted to conversations with S. Todadri, M. Barkeshli, D. Poilblanc, YC He, CM Jian,  J. Moore, and XL Qi,  the hospitality of Max Planck PKS, and the support of NSF DMR-1206515 and the David \& Lucile Packard Foundation.
AV thanks Xie Chen, Yuan Ming Lu and Max Metlitski for insightful discussions, and NSF-DMR 1206728 and the Templeton Foundation for support.

\bibliography{MI_set}

\appendix

\section{Existence of at least one symmetric MES}
\label{sec:sym_mes}
	A key technique used in the main text was to measure the 1D-SPT invariants associated with the local symmetries $G$ (for example $\gamma_x, S_x$), which must alternate between entanglement cuts when there is a projective representation per unit length.
However, the 1D-SPT invariant can only be defined if the MES is $G$-symmetric. 
One might worry that \emph{none} of the MES are invariant under $G$, but  instead are all permuted.
Let us clarify why  we can generally assume that there was at least one MES left invariant under the onsite $G$. Intuitively, it is the `vacuum' topological sector, but clearly this identification is slightly ambiguous when the unit cell is doubled.

For a connected continuous symmetry it is obvious, as there are no non-trivial representations from a connected continuous group to permutations, so all the MES are $G$-symmetric.

In the case when $G$ is time-reversal, for the double-semion model there should indeed  be a $\mathcal{T}$-symmetric MES. Referring to \ref{fig:mes}b), the only permutation of order two which exchanges the $s, s'$ edges is the reflection shown, which leaves two of the MES invariant.

In general, consider a very large cylinder which is capped off at the ends to form (topologically) a sphere. 
 On the sphere there is no topologically protected degeneracy, so if the sphere transforms with a linear representation under the symmetry (for example, it should have an even number of S = 1/2 sites), there should be some microscopic realization of the end-caps which results in a non-degenerate, $G$-symmetric low energy state. 
See Sec.~\ref{sec:exotic_T} for an exception when translation is realized in an exotic fashion.
Since we left the Hamiltonian  unchanged in the bulk of the cylinder, it is straightforward to argue that deep in the cylinder the wave function of this $G$-symmetric state is identical to some particular MES: consequently this MES is $G$-symmetric.

Why can't  the bulk of the cylinder look like a superposition of MES, perhaps like $\ket{a} + \ket{b}$?
In the 1D picture, these states have long-range order, and it is straightforward to show that to exponential accuracy the states $\ket{a}, \ket{b}$ would be eigenstates individually, as expected from spontaneous symmetry breaking of a finite length chain.
Hence to exponential accuracy there would actually be a doublet of states at the same energy - counter to the expected non-degeneracy of the sphere.

\section{Operational definition of $\eta_b$}
\label{sec:eta_n}
	We defined the Berry phase $\eta_b$ for dragging an anyon $b$ around one unit cell to be
\begin{align}	
\eta_b = \frac{(T^{-1}_y T^{-1}_x T_y T_x)_b} { (T^{-1}_y T_y T^{-1}_x T_x)_b}.
\end{align}
This might seem ill-defined if the anyon is an object much larger than the unit cell; how do we know whether we have dragged it precisely around the unit cell?

	To make this procedure rigorous, let $b$ be in the vicinity of $r$, with its anti-particle $\bar{b}$ far a way, near $r'$. Add pinning potentials $V_{r / r'}$ to the translation invariant Hamiltonian $H_0$, $H = H_0 + V_r + V_{r'}$, chosen such that the the pair is now the \emph{unique} ground state. If $b$ carries internal degrees of freedom, like spin, a symmetry breaking field should be included in $V$ to remove the degeneracy.
	
	To drag $b$ around, we make the pinning potential time dependent, $V_r(t)$, and require that 
\begin{align}	
V_r(1) &= T_x V_r(0) T_x^{-1}, \\
V_r(2) &= T_y T_x V_r(0) T_x^{-1} T_y^{-1}, \\
V_r(3) &= T_x^{-1} T_y T_x V_r(0) T_x^{-1} T_y^{-1} T_x,\\
V_r(4) &= V_r(0)
\end{align}
chosen such that \emph{there is a unique ground state} throughout the time evolution, and hence no level crossings.
See Sec.~\ref{sec:exotic_T} for an obstruction when translation is realized in an exotic fashion which forces level crossings.
For notational simplicity, we have marked four special points chosen at $t_n = n$, though this could of course be generalized as required to maintain adiabaticity.

We compute the adiabatic phase for the above cyclic process $(T^{-1}_y T^{-1}_x T_y T_x)_b$, which has a non-universal local contribution, and a topological contribution.
To separate out the topological contribution, repeat the above measurement, but re-arrange the above time dependence of $V_r(t)$:
\begin{align}	
\tilde{V}_r(t) = \begin{cases} V_r(t), & t\in [0, 1) \\
T_y^{-1} V_r(t + 1) T_y & t\in [1, 2) \\
T_x^{-1} V_r(t - 1) T_x & t\in [2, 3) \\
V_r(t) & t\in [3, 4) \\
\end{cases}
\end{align}
under which the particle acquires a phase $(T^{-1}_y T_y T^{-1}_x T_x)_b$.
Using $T_{x/y}$ symmetry, we see that each segment is locally identical to the earlier version, so by taking a ratio the local part cancels, giving a robust phase $\eta_b$.

Note that our procedure remains  well defined even if the pinned anyon is far larger than the unit cell, and we don't need to make any assumption that the particle travels in a `line' during the segments $t \in [n, n+1)$. Indeed, the potential can even be chosen to drag the anyon around other distant unit cells in the intervening periodic, as the resulting phases will still be canceled by the re-arranged version.

\section{Breakdown of the argument for certain exotic realizations of translation symmetry}
\label{sec:exotic_T}
	In certain models the anyon types are permuted by translations. For example, there are  Z$_2$ lattice models in which translation acts as an $e-m$ duality, $T_{x/y} e = m$.\cite{Kitaev2006, Bombin2010, You2012}
While this is impossible for double-semion topological order, the most general form of our no-go argument can be manifestly violated in these cases. 
In fact, CM. Jian has brought to our attention  a lattice model in which each unit-cell transforms \emph{projectively} under an onsite symmetry $G = \mathbb{Z}_2 \times \mathbb{Z}_2$, yet there is no anyonic excitation in the theory which carries a projective representation of $\mathbb{Z}_2 \times \mathbb{Z}_2$, counter to our argument.\cite{CMJian}
Here we briefly explain which assumptions of this work fail for this case.

	When $e/m$ are exchanged by translation, the ground state degeneracy on an infinite cylinder / torus is no longer given by the number of anyons $\gsd = 4$.
On an odd circumference cylinder, when an $e$-particle is dragged around the cylinder it is turned into an $m$-particle, and visa versa, so the process $\F_y^{e/m}$ is ill-defined. However, the processes  $(\F_y^{e})^2, (\F_y^{m})^2, \F_y^{f}$ remain well defined.
Restricting the algebra of Eq. 2) to these operations, we see that they are degenerate and can be realized in a $2$-dimensional ground state manifold. 
Hence the protection of the 4-fold degeneracy is lost, and is broken down to 2-fold.

One can still ask if the remaining 2-states are related by the translation $T_y$.
Again, the key technique used to demonstrate this in the main text was to measure the 1D-SPT invariant associated with the symmetry $G$, which must alternate between entanglement cuts when there is a projective representation per unit cell.
However, the 1D-SPT invariant can only be defined if the MES is $G$-symmetric. 
One might worry that \emph{none} of the MES are invariant under $G$, but  instead are all permuted.
This is what occurs for the counter example; the two MES are exchanged by $G$ and left invariant under translation!

The reason is that when $e$ and $m$ are permuted by translation, counter to the arguments in Sec.~\ref{sec:sym_mes} \emph{the two-fold degeneracy remains protected} by $G$ regardless of how the cylinder is capped off.
This phenomena has been understood in the framework of extrinsic  defects, where is it shown they carry a non-trivial quantum dimension of $\sqrt{2}$.\cite{BarkeshliJianQi2012}
There is now nothing which forbids an action of $G$ which exchanges the two ground states of the sphere, which is indeed what occurs in the counter example.

If there is no other mechanism which can  protect a 2-fold degeneracy of a capped off cylinder, we believe our line of argument remains true for discrete projective symmetries so long as the anyons are un-permuted by translation.

 	In addition, our operational definition of the invariant $\eta_b = \frac{(T^{-1}_y T^{-1}_x T_y T_x)_b} { (T^{-1}_y T_y T^{-1}_x T_x)_b} $ presumably breaks down as well.
Operationally, one must find a pinning potential which leads to a \emph{non-generate} ground state with a $b$-anyon pinned in the region, so that we may adiabatically drag $b$ around the unit cell. But if $b$ is transformed into another anyon, $T_x b \neq b$, presumably it is impossible to drag $b$ along $T_x$, since this would change the superselection sector. 
Indeed, this is exactly what happens in the Wen plaquette model, where $e$ particles are restricted to  odd plaquettes while $m$ particles are restricted to even plaquettes.\cite{Bombin2010, You2012}
As one attempts to drag the particle along $T_x$ with a pinning potential, there will be a level crossing where adiabatic transport breaks down.

\section{Absence of Kramers doublets in the double semion topological order}
\label{Appendix:Kramers}
Consider the double semion topological order $\{1, \,s, b,\, s'=bs \}$ with only time reversal symmetry $\T$. We have previously noted that the only quasiparticle that does not change its statistics under time reversal is `b', and hence this is the only one that can potentially exhibit  Kramers degeneracy in the bulk. 
We show here that even this is not possible, `b' must always be a Kramers singlet. The key property is that `b' may be viewed as a bound state of `s' and `s' which are exchanged by $\T$. Performing $\T$ twice, leads to a double exchange which acquires a phase equal to the mutual statistics. Since they are mutual bosons, this phase factor is unity hence $\T^2=+1$. Let us show this a bit more rigorously now.

\begin{figure}[t]
\begin{centering}
	\includegraphics[width=\columnwidth]{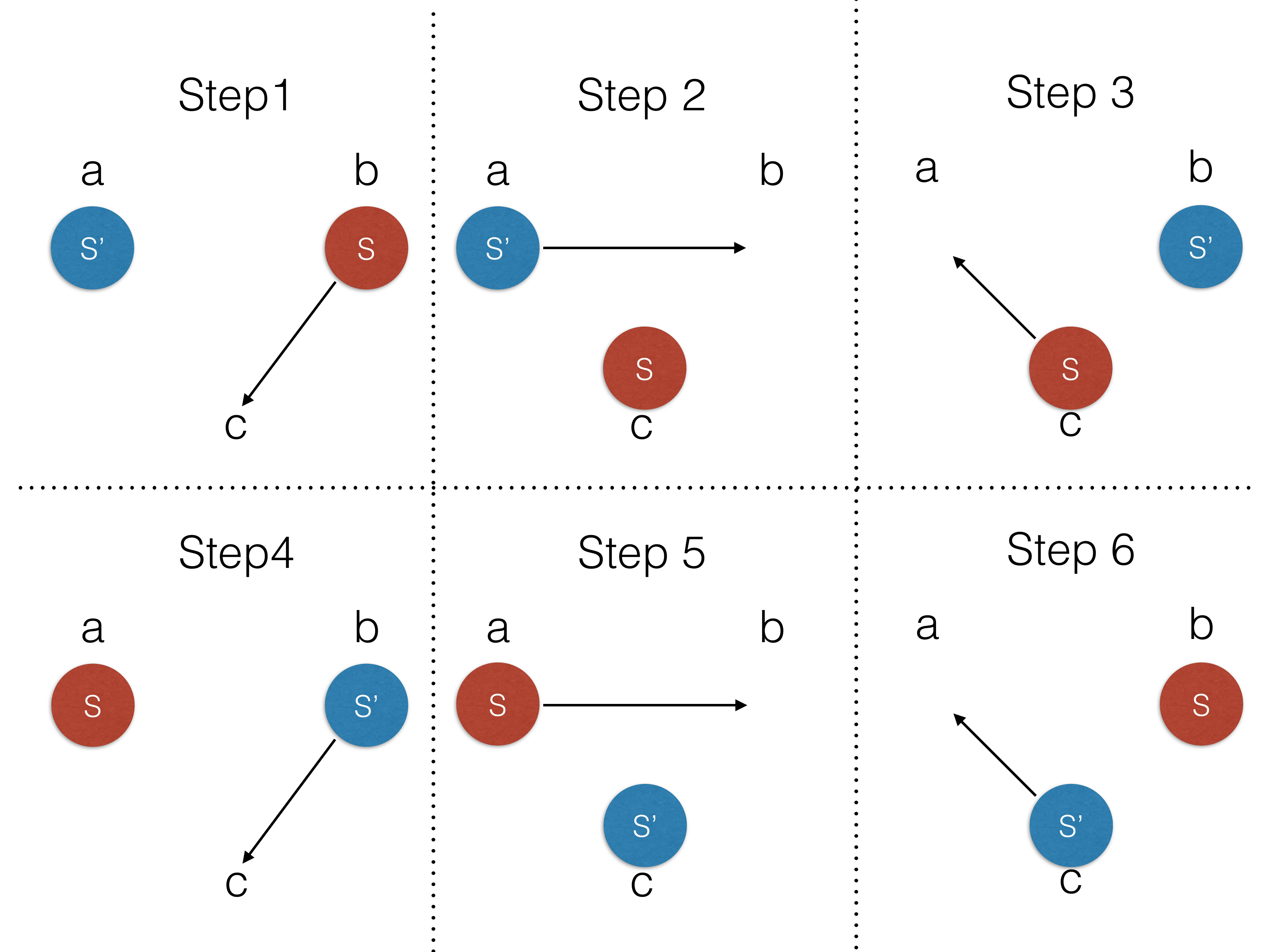}
	\caption{Time reversal symmetry acting on the `b' particle (composite of $ss'$) is the operator $X_2$ represented via steps 1-3 which ends up exchanging the $s\leftrightarrow s'$. The operator $\T X_2 \T^{-1}$ is shown in Steps 4-6. Combining them involves braiding one particle around the other and fixes Kramers degeneracy.}
	\label{fig:LocalT}
\end{centering}
\end{figure}

This result follows from an application \cite{MLevinPC} of local time reversal as discussed in Ref.~\onlinecite{LevinStern}.  Consider applying time reversal to a system with an even number of $S=1/2$ moments. Overall $\T^2=+1$. However, if within this system there are a pair of anyons - e.g. a pair of $b$ particles that were created from the vacuum, we would like to investigate the effect of time reversal on each of them. The action of time reversal on the ground state is localized near the position of the anyon (since this is a gapped phase which is time reversal symmetric). Say, under $\T=X_1X_2$, we perform local spin rotations $X_i$ near the two anyons, that implement the effect of time reversal symmetry. Then, local time reversal symmetry is implemented by $\T_1 = \T X_2$, which performs the operation on anyon `1' alone. Note also this definition of local $\T$ requires that the anyon itself does not change character, otherwise the operators $X$ would reach between a pair of anyons, obstructing a local definition of time reversal. 

We now wish to consider $\T_1^2= \T X_2\T X_2 = \T X_2 \T^{-1}X_2$. First let us write out $X_2$, which follows from the fact that `b' is a composite of s-s'. Time reversal exchanges them. This can be implemented by the sequence of operations $$X_2=(s':c\rightarrow a) (s:a\rightarrow b)(s':b\rightarrow c)$$
where each operation induces a string operator that moves the corresponding anyon from one location to the other. This accomplishes the necessary switch. Now let us compute:
$$\T X_2 \T^{-1}=  (s:c\rightarrow a) (s':a\rightarrow b)(s:b\rightarrow c)$$
where we simply exchange the labels of the particles. Combining these together we have:

\begin{eqnarray*}
\T_1^2 &=& (s:c\rightarrow a) (s':a\rightarrow b)(s:b\rightarrow c)\,  \\
&& (s':c\rightarrow a) (s:a\rightarrow b)(s':b\rightarrow c)
\end{eqnarray*}

this sequence simply corresponds to a double exchange - i.e. taking one particle around another. The resulting phase is unity since the particles have mutual bosonic statistics and hence we have a Kramers singlet $\T_1^1=+1$. 

Note, if instead the particles had mutual semionic statistics, then their fusion product would be Kramers degenerate under time reversal symmetry. Indeed this occurs in the 2D Toric code, with fermions in a $\T$ symmetric topological superconductor band structure when the $e$ and $m$ are exchanged by $\T$. Their fusion product, the fermion must be Kramers degenerate - which indeed is a requirement to obtain a topological superconductor band structure to begin with (class $DIII$).
	
\section{Entanglement invariants for U(1)}

\label{app:u1_inv}
There are no U(1) projective representations associated with $G = U(1)$, $\mathcal{H}^2( U(1), U(1) ) = \{ 1\}$, so it is not immediately clear what entanglement invariant can play the role of $\gamma_x, S_x$.
In the presence of translation symmetry, the 1D-SPT classification actually has additional data, the `charge per unit cell,' and here we discuss how the fractional filling leaves its imprint on the entanglement spectrum.
Consider the Schmidt decomposition $\sum_\alpha s_\alpha \ket{\alpha}_{<x} \ket{\alpha}_{>x}$ about the cut $x$.
For an MES, each left Schmidt state $\alpha$ can be assigned a U(1) quantum number$\{ Q_\alpha \in \mathbb{Z} \}$ (we assume a a fundamental charge of 1).
Because the Schmidt states are semi-infinite, the \emph{total} charge to the left is generally ill-defined, but the relative charges $Q_\alpha - Q_\alpha'$ are well defined.
Following a detailed discussion in Ref.~\onlinecite{ZaletelMongPollmann2013}, we use the entanglement spectrum to define the charge polarization $\langle Q \rangle_x$,
\begin{align}
e^{2 \pi i \langle Q \rangle_x} \equiv  e^{ 2 \pi i \sum_\alpha s_\alpha^2 Q_\alpha }.
\end{align}
$\langle Q \rangle_x$ is well defined modulo 1. 
For a featureless cylinder at filling $\frac{p}{q}$ per unit cell and circumference $L_y$, it is straightforward to prove
\begin{align}
\langle Q \rangle_{x+1} = \langle Q \rangle_x + L_y \frac{p}{q}.
\end{align}
This immediately proves a $q$-fold degeneracy for $L_y = \mathbb{Z} q + 1$.

While the behavior of $\langle Q \rangle_x$ requires a $q$-fold degeneracy, $\langle Q \rangle_x$ isn't itself a 1D SPT invariant so needn't be quantized.
However, additional symmetries can quantize  $\langle Q \rangle_x$.
For example, if the symmetry group is $U(1)  \rtimes \mathbb{Z}_2$ (bosons with particle-hole conjugation),  or  $U(1)  \times \mathbb{Z}^T_2$ (time-reversal acting on the spin $S^z$)  then for a symmetric state
\begin{align}
\langle Q \rangle_x  = 0 \mbox{ or } \frac{1}{2}.
\end{align}
Another example is a 180-degree  bond-centered spatial rotation we call `$C$'. 
Suppose there is an entanglement cut at  $x=0$ preserved under $C$.
With a slight change of notation, let $\langle Q \rangle_\psi$  denote the charge polarization in MES $\psi$ about $x = 0$.
Then
\begin{align}
\langle Q \rangle_{C \psi} &= - \langle Q \rangle_{\psi}	\quad \mbox{(mod 1)} \\
\langle Q \rangle_{T_x \psi} &= \langle Q \rangle_{\psi} + 1/q \quad \mbox{(mod 1)} \\
C T_x C  &= T_x^{-1}
\end{align}
For odd $q$, this requires either $\langle Q \rangle_\psi = p/q$, or $p/q  + \tfrac{1}{2}$. For even $q$, either $\langle Q \rangle_\psi = p/q$, or $(p + 1/2) / q$.

\end{document}